\documentclass[referee]{raa}            

\usepackage{graphicx,times}             
\usepackage{natbib}

\usepackage{amssymb,amsmath}

\usepackage[a4paper=true,dvipdfm=true,pagebackref=true]{hyperref}
\hypersetup{colorlinks = true, linkcolor = green, anchorcolor = red, citecolor = blue, filecolor = red, pagecolor = red, urlcolor = red}

\begin{document}

	\title{Distances to  18 Dwarf Galaxies from  the Arecibo  Survey\,$^*$
		\footnotetext{$*$ Based on observations with the NASA/ESA Hubble Space Telescope, obtained at the
			Space Telescope Science Institute,
			which is operated by AURA, Inc. under contract \textnumero\, NAS5-26555. These observations
			are associated with proposals 13750, 15243.}
	}
	\setcounter{page}{1}          
	\author{N.A.~Tikhonov\inst{1}, O.A.~Galazutdinova\inst{1}}
	
	\institute{Special Astrophysical Observatory, Nizhnij Arkhyz, Karachai-Cherkessian Republic, 
		Russia 369167; {\it ntik@sao.ru}\\
			\vs\no
		{\small Received~~{ 7, 2019  August; accepted~~ September 16, 2019}}}
	
	\abstract{Based on the archival Hubble Space Telescope images, we have performed stellar photometry for 18  dwarf  galaxies.   Branches of  young  and  old  stars  are  seen on  the  constructed  Hertzsprung-Russell diagrams.   Using  the  photometry  of  red  giants  and  applying  the  TRGB  method,  we  have  determined accurate  distances  for  all  18  galaxies  for  the first  time.   The  galaxies  AGC  238890  and  AGC  747826 have minimum ($D$\,=\,5.1\,Mpc) and maximum ($D$\,=\,12.0\,Mpc) distances, respectively.  The distances to the remaining galaxies lie within this range.  Low-metallicity galaxies have been identified by measuring the  color  indices  of  the  red  giant  branch:   AGC  102728,  AGC  198691,  AGC  205590,  AGC  223231, AGC  731921,  and  AGC  747826. We  have  determined  the  distance  to  AGC  198691  with  a  record low  metallicity.    Since  AGC  223254,  AGC  229053,  AGC  229379,  AGC  238890,  AGC  731921,  and AGC 742601 are projected onto the Virgo cluster of galaxies, the distances estimated by us together with the velocities of these galaxies measured previously at Arecibo can be used to refine the effect of galaxy infall to the Virgo cluster.
	}
	
	\authorrunning{N.A.~Tikhonov, O.A.~Galazutdinova}            
	\titlerunning{Distances to  18 Dwarf Galaxies from  the Arecibo  Survey}  
	
	\maketitle
	
	\section{INTRODUCTION}           
	\label{sect:intro}
	
A catalog of almost 16 000 objects, for which the coordinates,  H\,I fluxes,
  radial  velocities,  and  H\,I line widths were measured, was produced
 while conducting  the  ALFALFA  survey \citep{Giovanelli_etal_2005,Haynes_etal_2011}
  at  the  Arecibo  radio  telescope. In addition,  these  radio  sources
  were  identified  with optical counterparts from SDSS (Sloan Digital SkySurvey).
   Most  objects  of  the  catalog  turned  out  to be extragalactic sources,
 many of which are identified as  dwarf galaxies.  Some objects,  probable
 galaxies,  were  visible  in  the  radio  band,  but  were  absent in
  optical  sky  surveys.   These  were  assumed  to  be the so-called dark
 galaxies,  i.e.,  galaxies where star formation has not yet begun or
 proceeds very slowly. Such  galaxies  have  a  very  low  surface
  brightness and, therefore, are absent in the optical surveys, but they
 are  detected  with  confidence  at  the  radio  telescope \citep{Janowiecki_etal_2015}.
 Radio observations make it possible to find new dwarf galaxies in the
 vicinity of galaxy groups \citep{Cannon_etal_2011}, wich can change the forms
 of the luminocity functions of galaxy groups in the areas of their low mass
 members. The dwarf galaxies containing hydrogen and located far from
 neighboring galaxies arouse  special  interest. The  evolution  in such
 galaxies occurs without any external influence, and this make it possible
 to study the reasons that trigger star formation processis in these galaxie.

 However,   the  radio observations  alone  are  not  enough to  study  the  nature
  of  the  galaxies. To  calculate the  galaxy  masses  or  to  determine
  the  existence  of neighbors,  we  need  to  know  accurate  distances
  to the galaxies by attracting optical observations for this purpose.
 The TRGB method \citep{Lee_etal_1993}, which is based on measuring the
 position of the tip of the red giant branch, is most accurate and popular.
 We used this method to determine the distances to 18 galaxies whose images
 were obtained with the Hubble Space Telescope (HST) in 2015, but their
 distances were not determined.
\begin{figure}[h]
	\centerline{\includegraphics[angle=0, width=14cm,clip]{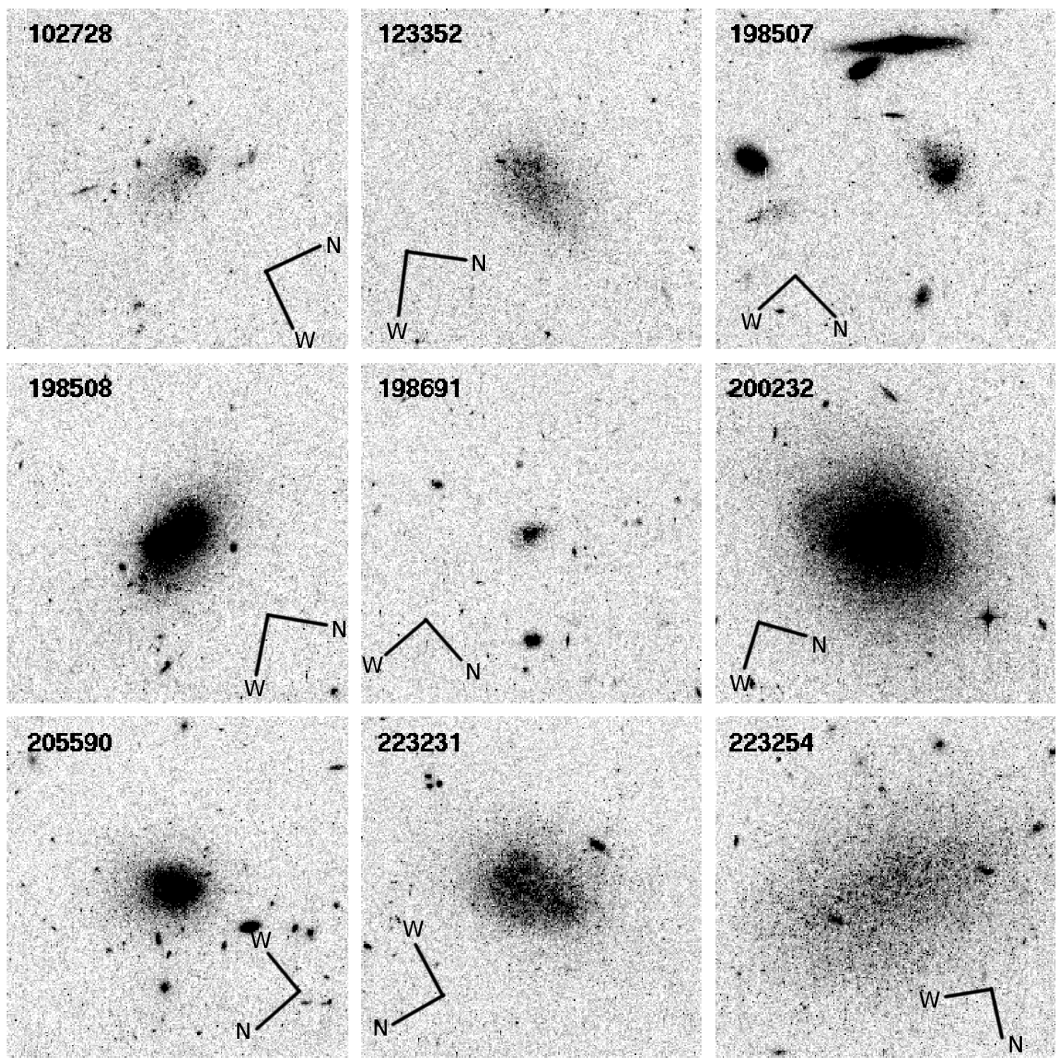}}
	\caption{HST ACS/WFC images of the galaxies. The sizes of each image are $1.0\arcmin \times 1.0\arcmin$. A large difference between the galaxy sizes is clearly seen.}\label{fig1}
\end{figure}
\setcounter{figure}{0}
\begin{figure}[h]
	\centerline{\includegraphics[angle=0, width=14cm,clip]{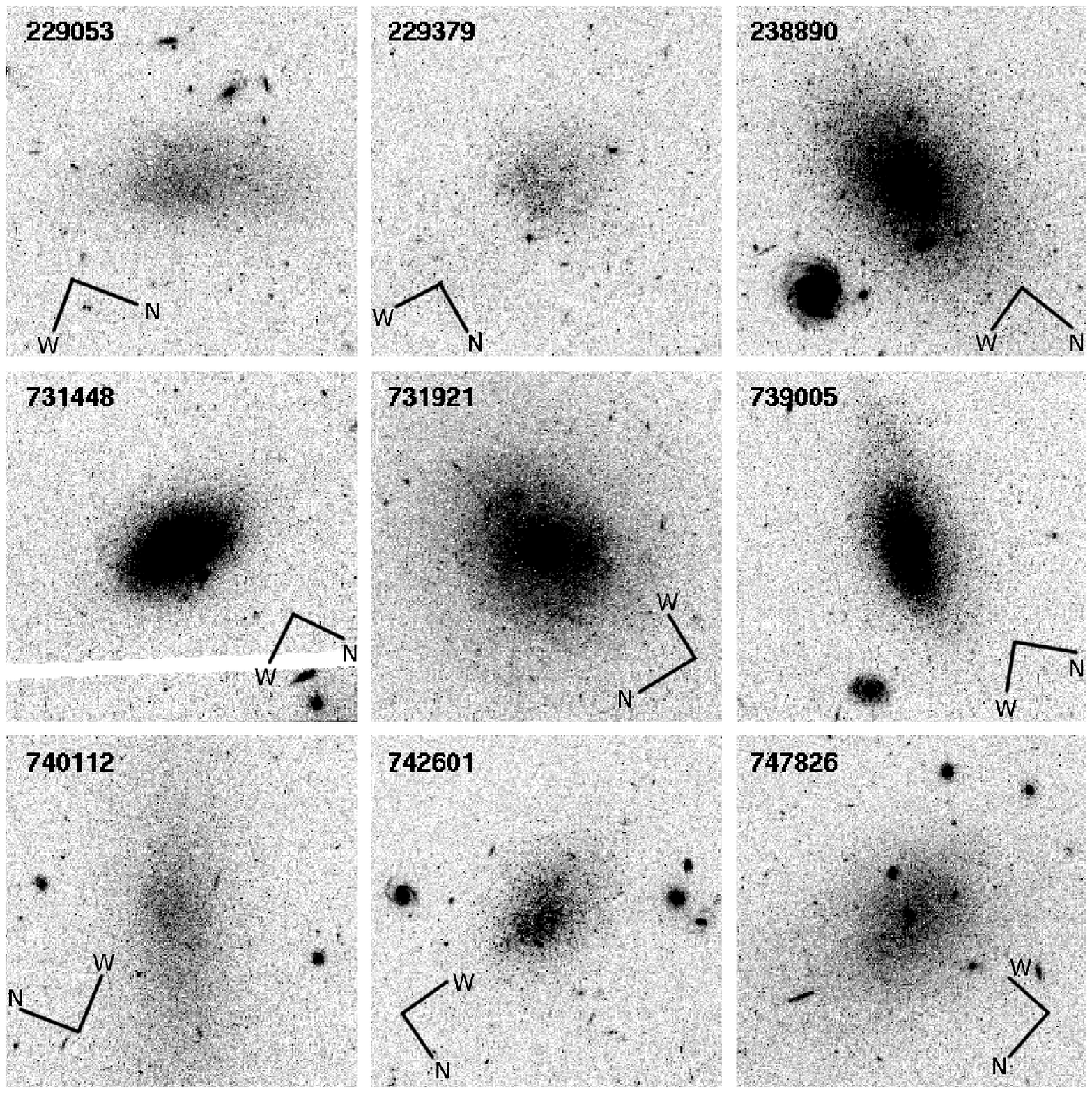}}
	\caption{(Contd.)}
\end{figure}

\section{STELLAR PHOTOMETRY}
HST  ACS/WFC  images  were  obtained  on  proposal  ID  13750  (J.  Cannon)  in  the  F814W  and F606W filters   with   exposure   times   of   2648\,s   and 2510\,s. Additional HST WFC3 images in the F814W and  F606W filters  with  exposure  times  of  18618\,s and  15018\,s  were  obtained  on  proposal  ID~15243\,(K.~McQuinn)  for  AGC\,198691,  which  turned  out to  be  a  dwarf  galaxy  with  a  very  low  luminosity ~\citep{Hirschauer_etal_2016}. 

The  averaged  images  of  the  galaxies  with  the F814W  and  F606W filters
  are  presented  in  Fig.~\ref{fig1}. All  images  are  presented  on  the
 same  scale,  which clearly shows a variety of linear sizes and masses of
 the investigated galaxies, even despite the difference in  the  distances
  to  these  galaxies. AGC\,198691, the smallest one among the  18  galaxies,
  has  a  size of  0.43~kpc,  while  AGC\,731921  is  almost  10~times larger,
 3.4~kpc.  The linear sizes of the galaxies were estimated  from  the
  distribution  of  red  giants  along their radius. The exponential
 distribution of red giants on a logarithmic scale is represented  by
 a  linear  function, which allows one to determine the limiting radius
 at which the distribution of red giants intersects with a horizontal
 background line consisting of  distant galaxies and CCD  noise.

 The  stellar  photometry  of  all  galaxies  was  performed  with  two
  software  packages:  DAOPHOT~II \citep{Stetson_1987,Stetson_1994}  and
  DOLPHOT~2.0  \citep{Dolphin_2016}\footnote{http://americano.dolphinsim.com/dolphot/dolphot.pdf}.
 The stellar photometry in DAOPHOT~II was  carried  out  in  a  standard
  way,  as  we  described previously \citep{Tikhonov_etal_2009}, while
 the calibrations were obtained on the basis of stellar photometry with
  different  detectors  and  at  different  telescopes \citep{Tikhonov_and_Galazutdinova_2009}.
 The obtained results of stellar  photometry were selected according to the
 CHI and SHARP parameters, which define the shape of the photometric profile
 of each measured star \citep{Stetson_1987}.  This allowed
 us to remove all diffuse objects  (star  clusters,  distant  or  compact
  galaxies) from the photometry tables, because the photometric profiles
  of  these  objects  differed  from  those  of  the isolated stars that
 we chose as standard ones.
 
 The DOLPHOT~2.0 package was used in accordance  with  Dolphin's  recommendations,  while  the photometry  procedure  consisted  of  bad  pixel  premasking, cosmic-ray particle hit removal, and further PSF photometry for the stars found in two filters. The selection of our list of stars by the CHI and SHARP image profile parameters was made in the same way as in DAOPHOT~II. 
 
 \begin{figure}[h]
 	\centerline{\includegraphics[angle=0, width=12cm,clip]{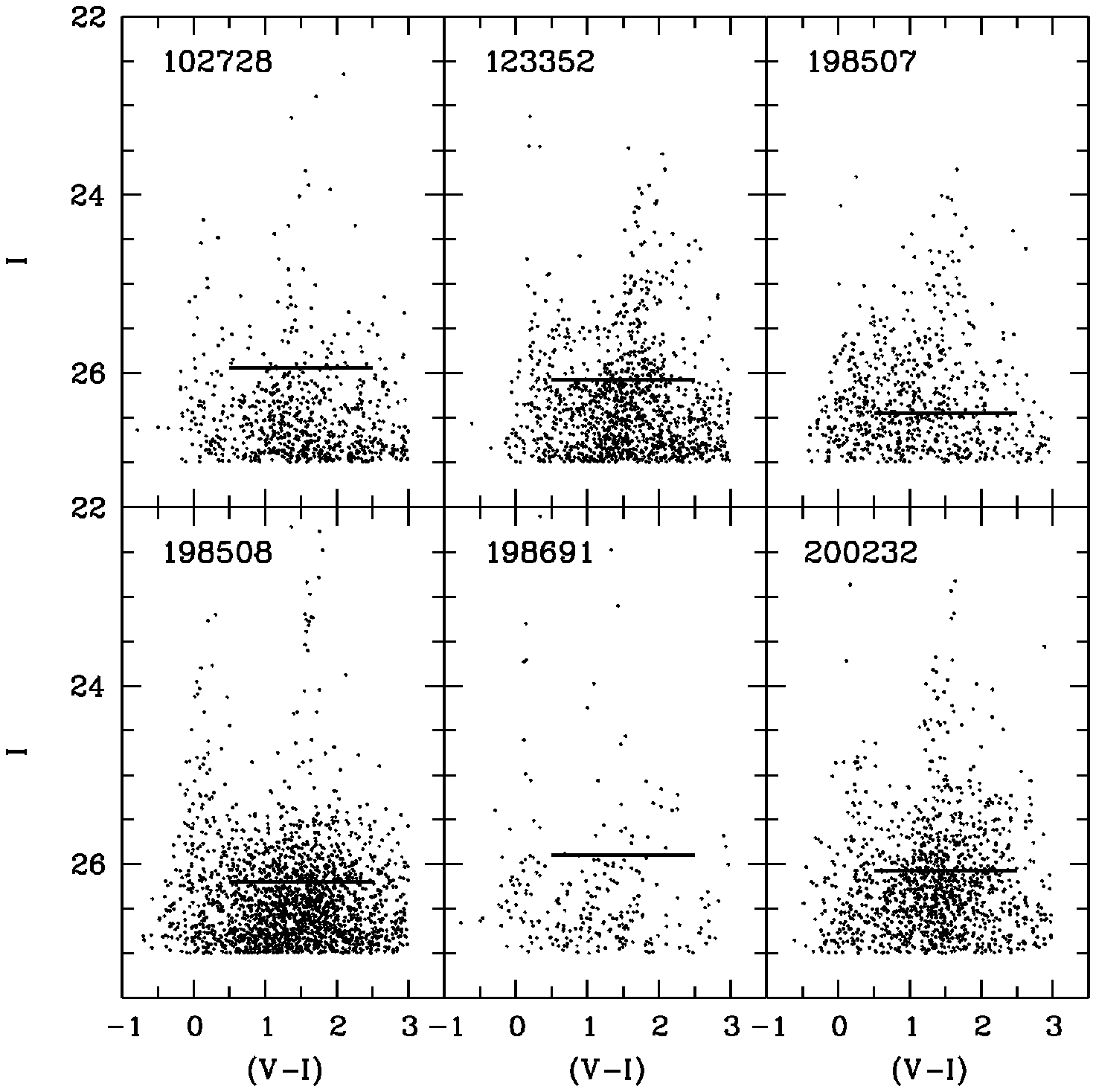}}
 	\caption{CM diagrams for six galaxies from the list. The horizontal lines mark the positions of the TRGB jumps. For the galaxies where the positions of the TRGB jumps raise doubts, Fig.~3 presents the luminosity functions with an additional selection of stars, whose details are described in the text.}\label{fig2}
 \end{figure}
 \setcounter{figure}{1}
 \begin{figure}[h]
 	\centerline{\includegraphics[angle=0, width=12cm,clip]{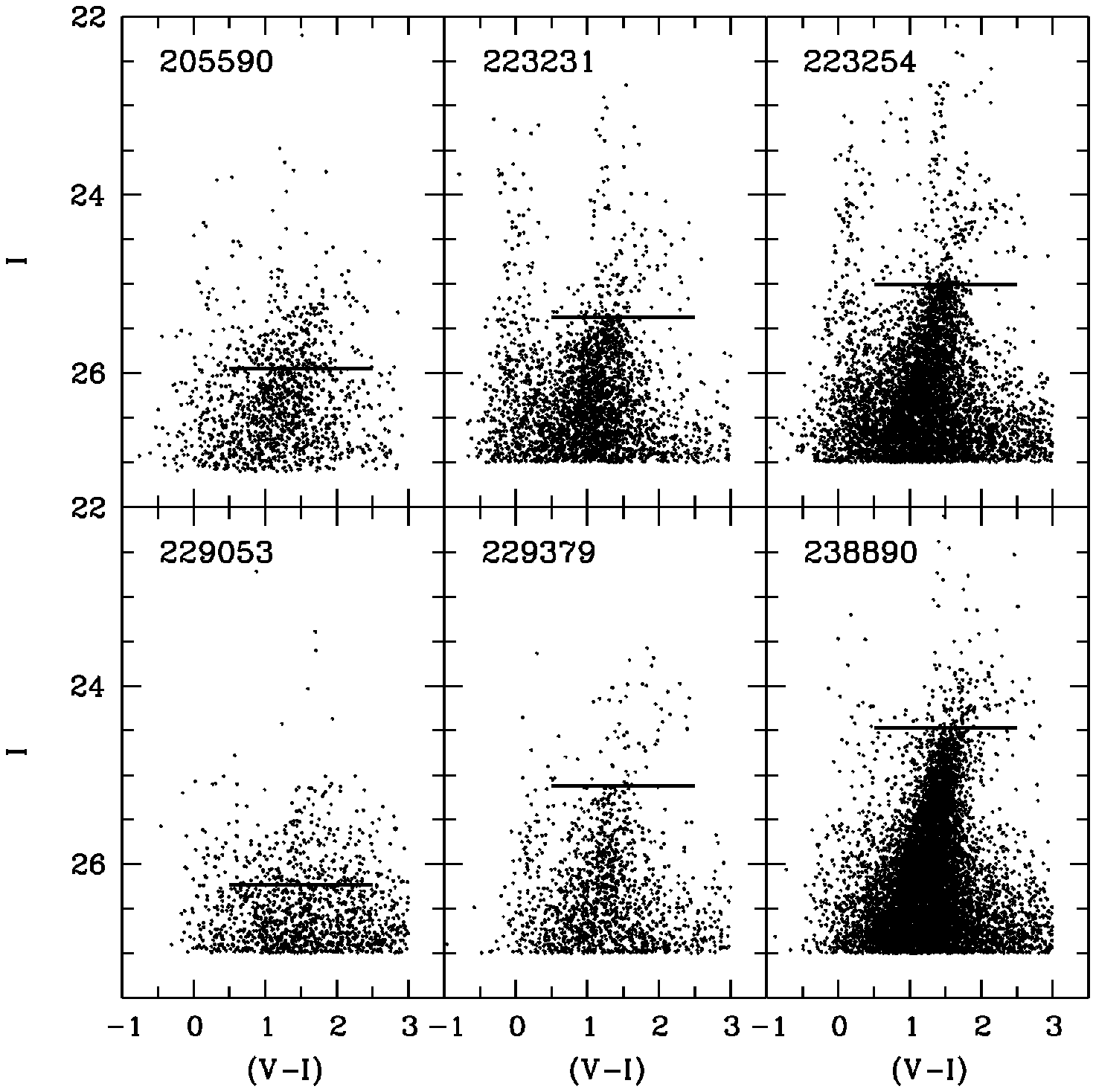}}
 	\caption{(Contd.)}
 \end{figure}
 
 \setcounter{figure}{1}
 \begin{figure}[h]
 	\centerline{\includegraphics[angle=0, width=12cm,clip]{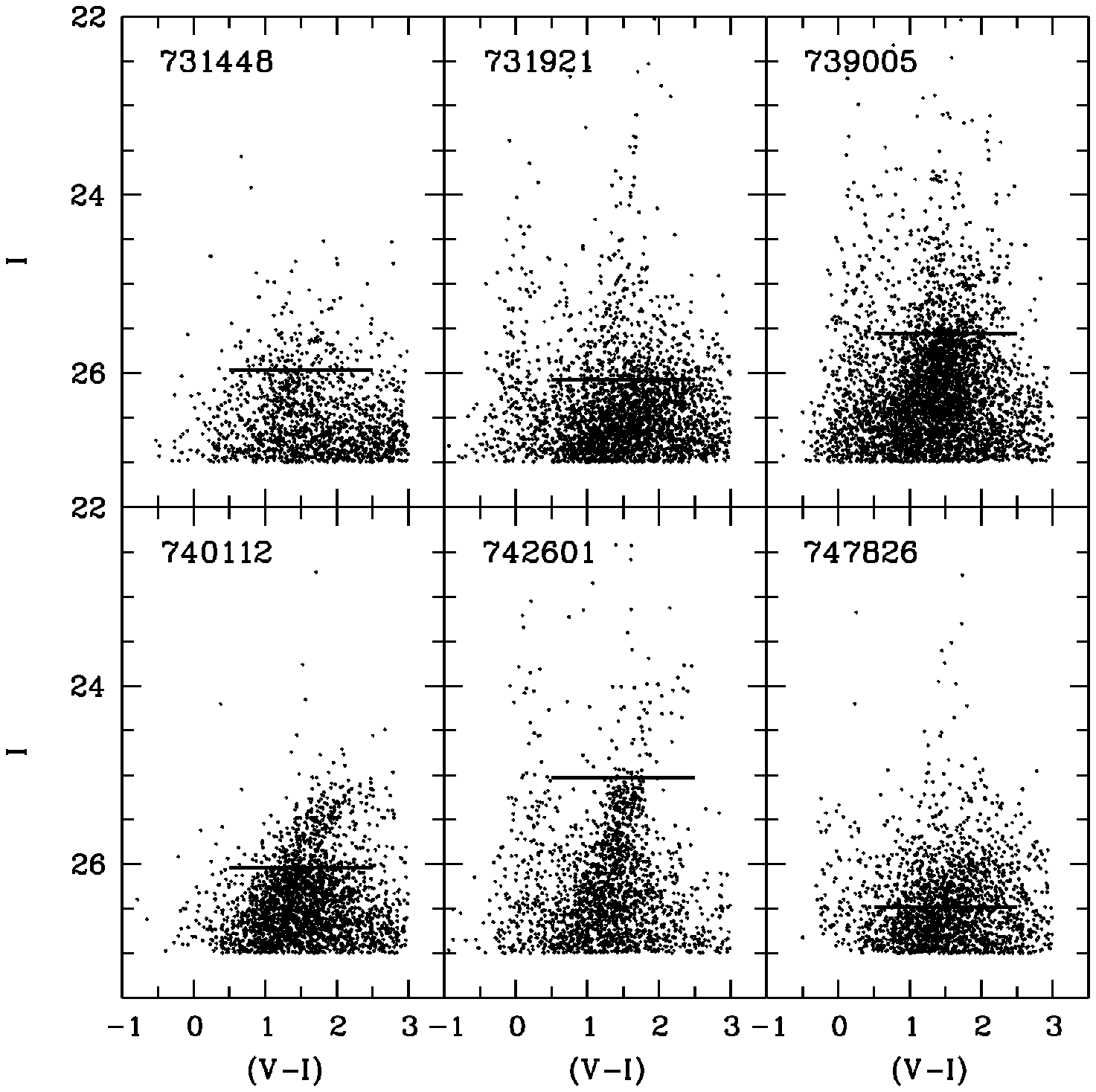}}
 	\caption{(Contd.)}
 \end{figure}
 The  principles  of  DOLPHOT  and  DAOPHOT photometry are the same,
  but there are some differences in using them.  For example, we used stars
 in the images of the investigated galaxies as PSF stars in DAOPHOT, while
 a library of PSF profiles was used in DOLPHOT. A difference between the
 results of the two software packages is seen when comparing the apparent distributions
 of very faint stars over the image field. Because of the charge transfer
 inefficiency and the existence of residual cosmic ray traces,
 DOLPHOT shows an excessive number of faint stars in the central region
 of the field instead of their even distribution, while the distribution
 of stars in DAOPHOT is closer to the real one. However, the problem of
 choosing PSF stars arises in DAOPHOT because of the appearance of
 ``tails'' due to the charge transfer inefficiency. Bearing in mind the
 pluses and minuses of the two software packages, we used them both by
 comparing the results obtained. Since we used stars brighter than the
photometric limit by two or more magnitudes for our measurements
 of the TRGB jumps and stellar metallicity, both methods yielded similar
 results and no significant differences between them were found.

The Hertzsprung--Russell diagrams (color--magnitude (CM) diagrams) for the 18 galaxies constructed from our stellar photometry are presented in Fig.~\ref{fig2}. The horizontal lines mark the TRGB jumps, i.e., the TRGB positions that we used to determine the distances to the galaxies.

\section{DISTANCE MEASUREMENTS}

The  intensive  use  of  red giants  to  determine  the distances to galaxies
 by the TRGB method has begun after the study performed by \citet{Lee_etal_1993},
  and by now accurate  distances  to several  hundred  galaxies have been
  measured  by  this  method. As  any  distance measurement  method, the
  TRGB  method  has  its difficulties  of  application. A  small  number  of
  red giants  in  a  galaxy  on  the  constructed  CM  diagram or an insufficient
 photometric limit of its images lead to  great  uncertainties  in  measuring
  the  position  of the  TRGB  jump  and,  hence,  to  a  low  accuracy  in
 measuring the distance to the galaxy.  Furthermore, the  fact  that  a  charge
  transfer  inefficiency  appeared in  the  ACS/WFC in  the  time
  of  their work under cosmic radiation
 \citep{Anderson_and_Bedin_2010,Massey_etal_2010,Tikhonov_and_Galazutdinova_2016},
  which becomes progressively  larger from year to year,  should be taken into
 account.  The  central  part  of  the  ACS/WFC  became  virtually
  unsuitable  for  accurate  photometric  measurements  due  to  this  effect.
   Bearing this  in  mind,  we  did  not  use  the  central  part  of  the ACS/WFC
 field with $1200$\,pix$<Y<3000$\,pix where this  was  possible.   Apart  from  red  giants,
  there  are brighter asymptotic giant branch (AGB) stars in each galaxy, which
 smear the TRGB jump on the CM diagram and make it difficult to measure the distance.
 Since  the  red  giants  and  AGB  stars  have  different number  density  gradients
  along  the  galactic  radius \citep{Tikhonov_2005,Tikhonov_2006},  we  can  reduce
  the  number of  AGB  stars  in  the  sample  using  only  the  galactic periphery
  for  our  measurements,  which  allows  the position  of  the  TRGB  jump  to  be
  measured  more accurately.

\begin{figure}[h]
	\centerline{\includegraphics[angle=0, width=10cm,clip]{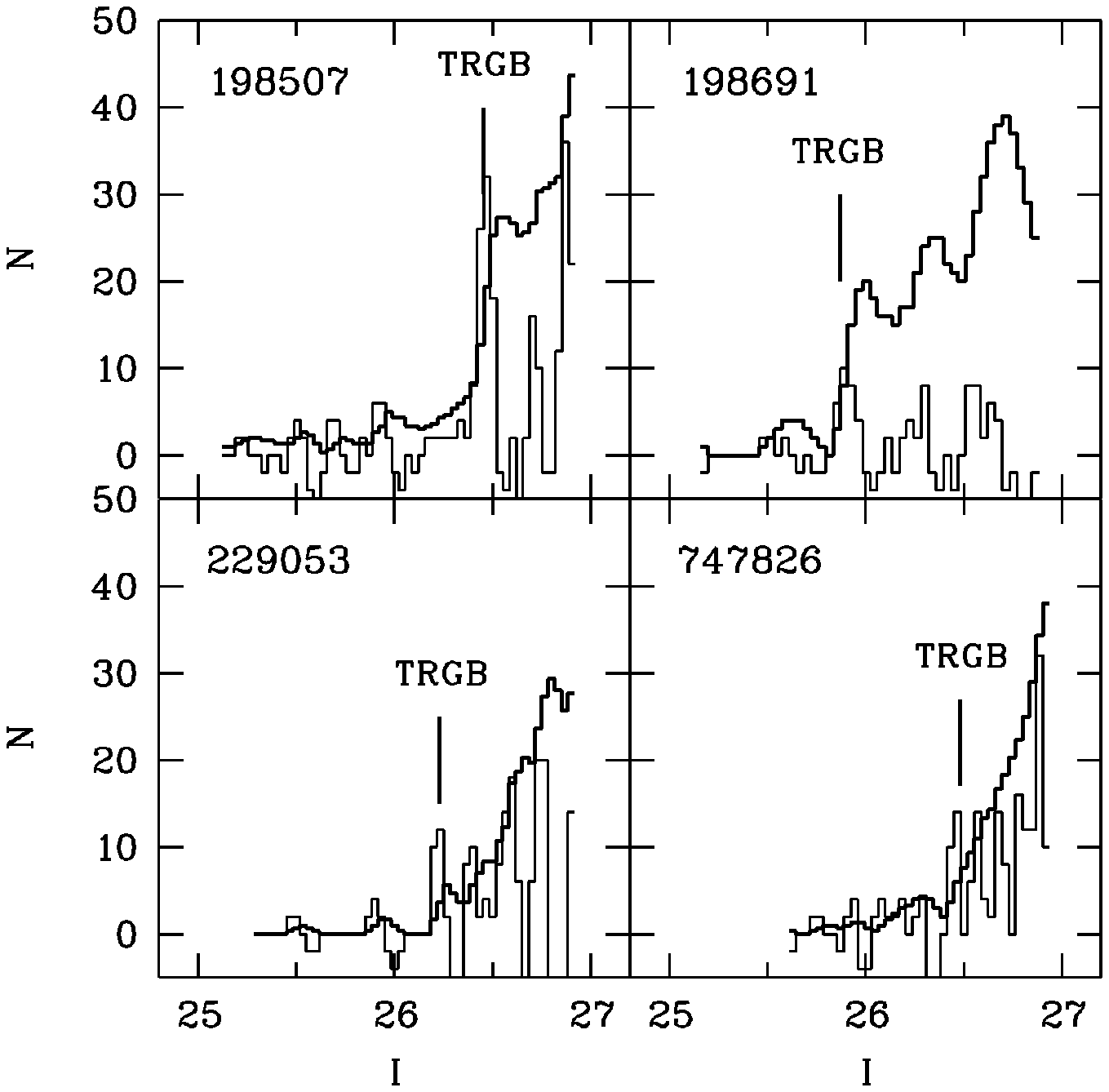}}
	\caption{Stellar luminosity functions of the galaxies for which the position of the TRGB jump is difficult to determine on the CM diagram. To construct these luminosity functions, we removed the stars in the central regions of the galaxies, where most AGB stars are located, from our sample of stars.}\label{fig3}
\end{figure}

For most galaxies, the TRGB is seen on the CM diagrams  quite  clearly,   but  for  several  galaxies  this position  is  not  obvious.    For  these  galaxies  Fig.~\ref{fig3} presents  the  luminosity  functions  of  red  giants  and AGB  stars.    We  used  samples  of  stars  on  the  periphery of the galaxies to construct these luminosity functions.  There are no bright supergiants in such a sample, and the number of undesirable AGB stars is reduced significantly.  The thin line on the luminosity function of each  galaxy (Fig.~\ref{fig3})  indicates  the Sobel function \citep{Madore_and_Fridman_1995} whose maxima correspond to abrupt changes in the number of stars, which is observed at the RGB boundary and is defined as the TRGB jump. 

In the galaxies where the red giants is clearly seen, these selection  by  CHI,  SH,
  and  CCD  coordinates $Y>3000$ pix was  sufficient  to  determine  the  TRGB
  jump. These restrictuon turned out to be insufficient for the remote
  galaxies,  and  additional  selections  were  applied.    The  stars  in
  the  central  parts  of  the  galaxies were removed, which increased the
 percentage of red  giants  in  such  a  sample.   In  addition,  we  made
 a selection  by  color,  usually $1.0<(V-I)< 1.7$, which  eliminated  the
  main-sequence  stars  and  the AGB  stars  with  a  large  color  index
  in  the  sample. Since  we  studied  the  dwarf galaxies  where  sparsely
 populated RGBs are visible, a difference by ten stars per  each  bin of the
 luminosity function was  enough for  the  TRGB  jump  to  be  clearly  seen.
   The  actual position  of  the  TRGB  jump  was  checked  by  examining the
 available luminosity function in logarithmic coordinates  of  the  number
  of  stars. As  a  rule,  a break at the point of the TRGB jump is seen on
 the luminosity function, which validated the choice.

When determining the distances, we measured the positions of the TRGB jumps as well as the TRGB colors $(V-I)_{\rm{TRGB}}$ and  RGB  colors $(V-I)_{-3.5}$ at $M_I = -3.5$. Using these quantities and  equations from \citep{Lee_etal_1993},  we  determined  the metallicities  of red giants and the distance moduli for the galaxies.

\begin{table}[h]
	\begin{center}
	\caption{Parameters of the 18 AGC galaxies} \label{Tab1}
	\scalebox{0.95}{
		\begin{tabular}{ccccccccccc} \hline
			N & AGC & $\alpha$  & $\delta$  & $I_{\rm{TRGB}}$  & $A_V$ & $(m-M)$ & [Fe/H] & $D$ & $\Delta{D}$ & $\Delta_{\rm{M87-galaxy}}$
			\\
			\hline
			01 & 102728 & 00 00 21.42 & +31 01 18.7 & 25.94 & 0.126 & 29.73 & -2.77 & 08.84 & 0.68 & 136\\
			02 & 123352 & 02 48 39.19 & +23 16 27.1 & 26.07 & 0.678 & 29.64 & -2.18 & 08.47 & 0.65 & 131\\
			03 & 198507 & 09 15 25.79 & +25 25 10.4 & 26.45 & 0.090 & 30.37 & -2.19 & 11.85 & 0.85 & 48\\
			04 & 198508 & 09 22 56.97 & +24 56 48.5 & 26.20 & 0.098 & 30.09 & -2.22 & 09.97 & 0.70 & 46\\
			05 & 198691 & 09 43 32.40 & +33 26 58.2 & 25.90 & 0.038 & 29.74 & -2.88 & 08.88 & 0.75 & 44\\
			06 & 200232 & 10 17 26.50 & +29 22 11.0 & 26.07 & 0.082 & 30.01 & -1.80 & 10.06 & 0.72 & 35\\
			07 & 205590 & 10 00 36.56 & +30 32 10.1 & 25.95 & 0.051 & 29.85 & -2.23 & 09.34 & 0.68 & 39\\
			08 & 223231 & 12 22 52.68 & +33 49 44.4 & 25.37 & 0.035 & 29.26 & -2.46 & 07.13 & 0.46 & 22\\
			09 & 223254 & 12 28 05.07 & +22 17 28.2 & 25.00 & 0.057 & 28.94 & -1.96 & 06.15 & 0.40 & 10\\
			10 & 229053 & 12 18 15.49 & +25 34 05.1 & 26.23 & 0.049 & 30.21 & -1.84 & 11.02 & 0.82 & 11\\
			11 & 229379 & 12 30 34.01 & +23 12 20.2 & 25.12 & 0.075 & 29.03 & -2.18 & 06.40 & 0.41 & 14\\
			12 & 238890 & 13 32 30.35 & +25 07 24.5 & 24.47 & 0.036 & 28.53 & -1.22 & 05.08 & 0.37 & 19\\
			13 & 731448 & 10 23 44.97 & +27 06 39.8 & 25.96 & 0.077 & 29.94 & -1.61 & 09.73 & 0.70 & 33\\
			14 & 731921 & 12 05 34.27 & +28 13 56.2 & 26.07 & 0.057 & 29.98 & -2.23 & 09.89 & 0.72 & 17\\
			15 & 739005 & 09 13 38.98 & +19 37 07.8 & 25.55 & 0.121 & 29.47 & -1.80 & 07.83 & 0.50 & 48\\
			16 & 740112 & 10 49 55.40 & +23 04 06.2 & 26.04 & 0.122 & 29.98 & -1.61 & 09.90 & 0.70 & 26\\
			17 & 742601 & 12 49 36.87 & +21 55 05.6 & 25.03 & 0.095 & 29.00 & -1.60 & 06.31 & 0.40 & 11\\
			18 & 747826 & 12 07 49.99 & +31 33 07.9 & 26.48 & 0.055 & 30.40 & -2.16 & 12.01 & 0.87 & 20\\
			\hline
	\end{tabular}}
\end{center}
\end{table}

The extinction toward each galaxy was taken from \citet{Schlafly_and_Finkbeiner_2011}.  Our results are presented  in  Table~\ref{Tab1},  where $\alpha$ and $\delta$ are  the  right  ascension and declination of each galaxy, $I_{\rm{TRGB}}$ is the position of the TRGB jump on the luminosity function in the $I$ band, $A_V$ is the extinction in the $V$ band in magnitudes, $(m-M)$ is the distance modulus, [Fe/H] is  the  metallicity  of  red  giants  on  the  galactic  periphery, $D$ is the distance to the galaxy in Mpc, $\Delta{D}$ is the external  distance  measurement  accuracy,  and $\Delta_{\rm{M87-galaxy}}$ is the angular distance (in degrees) from the  galaxy  to  M\,87,  the  central  galaxy  of  the  Virgo cluster. 

The accuracy of the distance measurement is individual for each galaxy. However, all galaxies can be arbitrarily divided into two groups by the distance measurement accuracy.   The first  group  includes  most  of  the  galaxies where  the  RGB  is  clearly  seen  and  the  position  of the  TRGB  jump  is  determined  with  an  accuracy  of  $0.^{\rm {m}}02$--$0.^{\rm {m}}03$.  For these galaxies, the internal accuracy  of  the  distance  is $0.2$~Mpc.   To  determine  the external  accuracy,  we  should  take  into  account  the accuracy of the TRGB method itself, which is $0.^{\rm {m}}1$. Given  the accuracy  of  other quantities,  the external accuracy  of  the  distances  for  such  galaxies  will  be $0.4$--$0.5$~Mpc.   For  the  galaxies  where  the  RGB  is seen more poorly (Fig.~\ref{fig3}), the accuracy of measuring the TRGB jump is $0.^{\rm {m}}04$--$0.^{\rm {m}}06$.  For these galaxies,  the  internal  accuracy  is $0.3$--$0.4$~Mpc,  while the external  accuracy  is  $0.7$--$0.8$~Mpc.   For each  galaxy Table~\ref{Tab1} gives the external distance measurement accuracy determined  from the width of the peak of the Sobel  function,  the  accuracy  of  our  photometry  for PSF  stars,  and  the  accuracy  of  the  TRGB  method itself.

\begin{figure}[h]
	\centerline{\includegraphics[angle=0, width=7cm,clip]{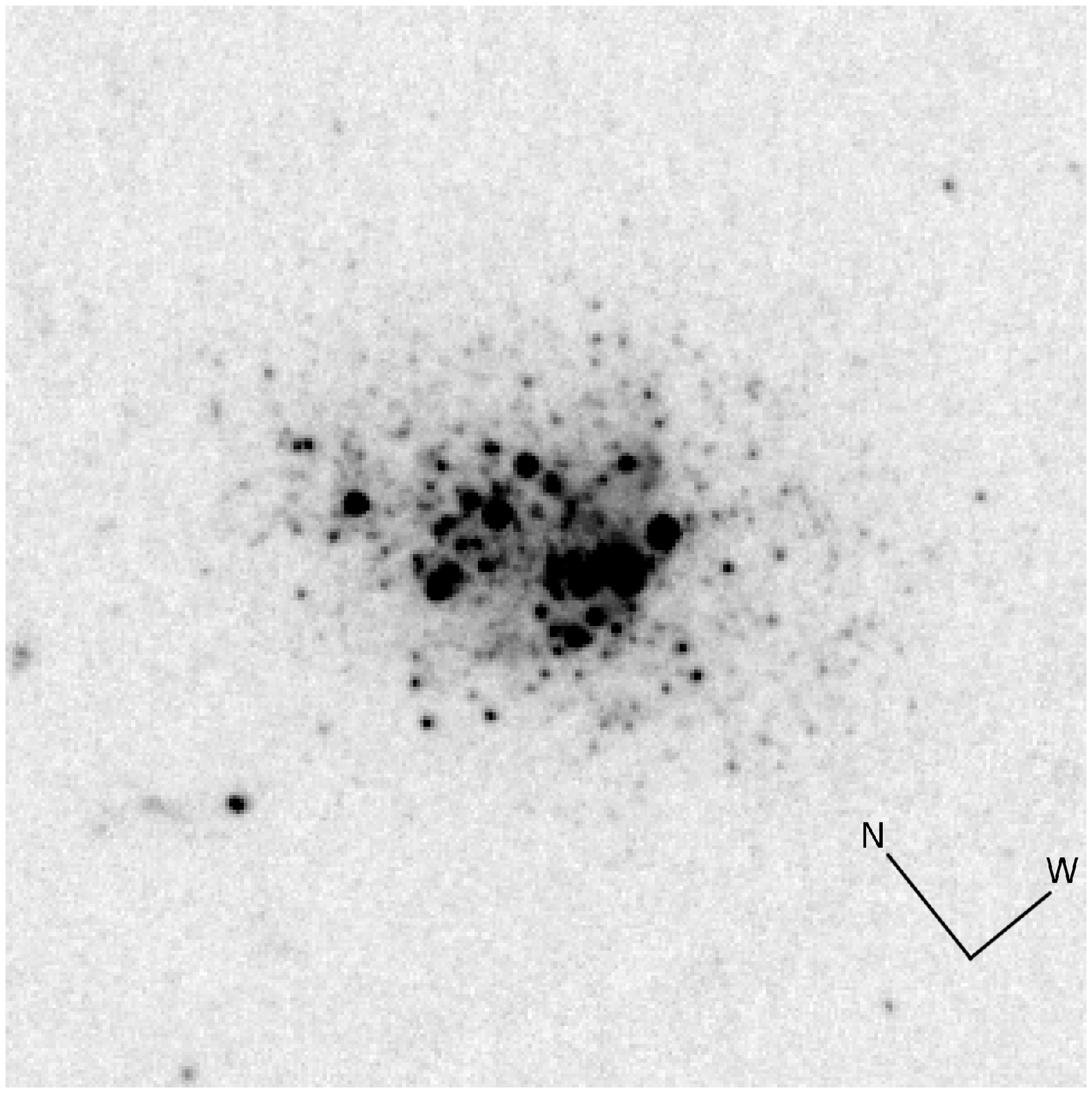}}
	\caption{WFC3 image of AGC198691 in the F606W filter. The image sizes are
		$15\arcsec \times 15\arcsec$.}\label{fig4}
\end{figure}

\begin{figure}[h]
	\centerline{\includegraphics[angle=0, width=12cm,clip]{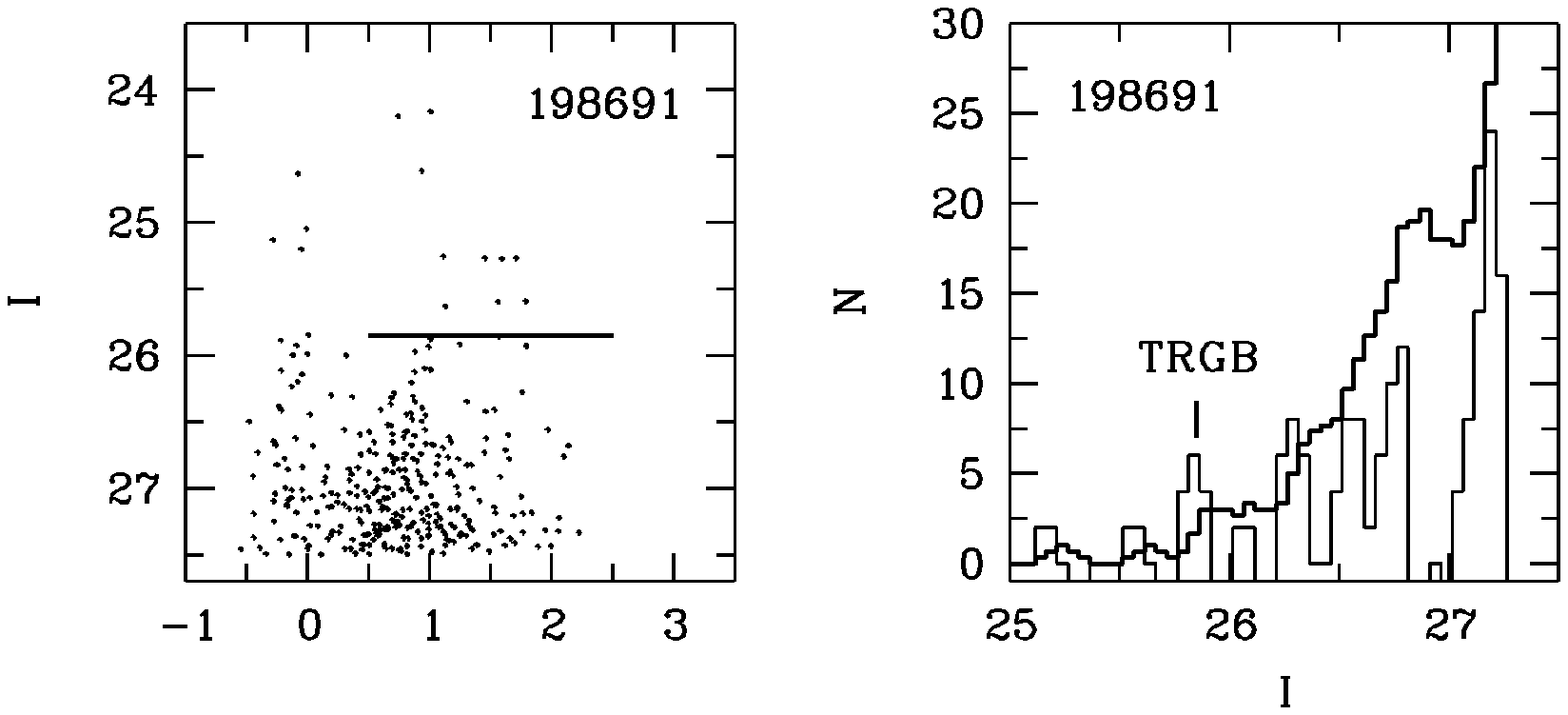}}
	\caption{CM diagram for AGC\,198691 stars from the WFC3 images and the luminosity function of AGB stars and red giants.}\label{fig5}
\end{figure}

For the galaxy AGC 198691, in which \citet{Hirschauer_etal_2016}  obtained  a  record  low  luminosity,  the applicants  of  the  observing  program  failed  to  measure the distance from the ACS images (ID~13750). Therefore, McQuinn obtained deeper WFC3 images for  it (ID~15243). We  processed  these  additional images~(Fig.~\ref{fig4}) using the above mentioned technique and presented the CM diagram and the luminosity function with the marked position of the TRGB jump in Fig.~\ref{fig5}. The distance estimated from the WFC3 images corresponds to the distance estimated from the ACS images. 

\begin{figure}[h]
	\centerline{\includegraphics[angle=0, width=10cm,clip]{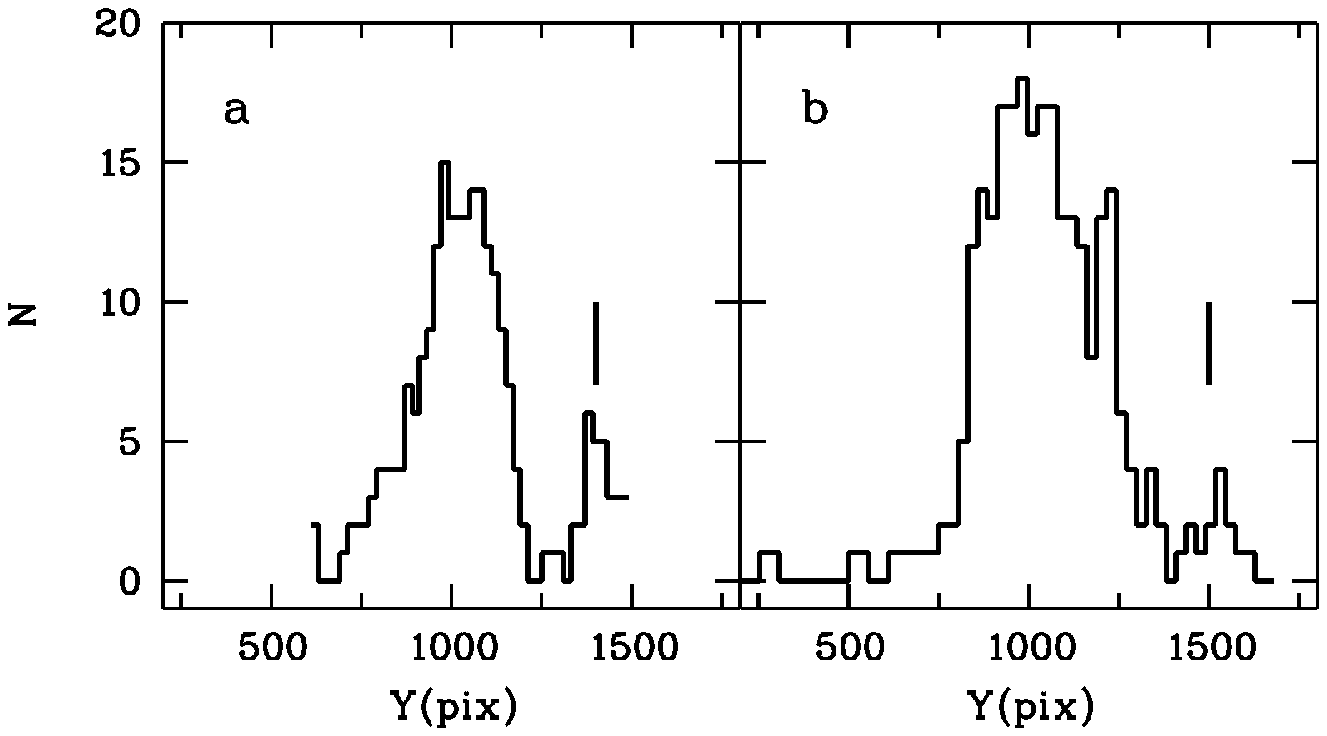}}
	\caption{The distribution of young (a) and old (b) stars along the major axis of AGC\,198691. The vertical line marks the position of the dwarf satellite near the main galaxy. The concentration centers of red giants and young stars are slightly shifted relative to each other due to the low statistics and the real asymmetry of the star-forming region.}\label{fig6}
\end{figure}

\section{CLOSE  NEIGHBORS}

The HST image size is $3.\arcmin5$. For a galaxy at a distance of 10~Mpc this corresponds to 10~kpc.  If there is a neighboring galaxy at a distance less than 5~kpc near an AGC galaxy, then it will be seen in the same image.    AGC\,198507  has  such  a  neighbor,  where a  dwarf  galaxy  that  may  be  called  AGC\,198507A (Fig.~\ref{fig1}) is seen at a distance of $30\arcsec$ (corresponding to 1.8~kpc). This galaxy contains few stars, but we were able to measure the position of the TRGB jump and to determine that the distance to this galaxy is equal, within the measurement error limits, to the distance to the main AGC\,198507.  Thus, these galaxies constitute a physical pair.  The asymmetry in the shapes of  these  galaxies  can  possibly  be  explained  by  their interaction.

Two  centers  are  observed  in  the  apparent  distribution  of  AGC\,739005  stars. This  is  particularly clearly  seen  in  the  distribution  of  young  stars,  red supergiants. Based  on  the  apparent  morphology, AGC\,739005  can  be  represented  as  two  galaxies spaced  0.73~kpc  apart,  one  of  which  is  a  symmetric Sph/Irr galaxy and the other one is irregular.  Since two  concentration  centers  of  stars  are  also  seen  in the  distribution  of  red  giants  (Fig.~\ref{fig6}),  one  of  which corresponds  to  AGC\,739005A,  this  satellite  cannot be  a  star-forming  region,  but  is  a  dwarf  irregular galaxy with young and old stellar populations and a very low stellar metallicity. 

AGC\,198508 has an approximately similar shape as AGC\,739005, where a star-forming region or a small galaxy is located  at  the  edge  of  the  galaxy.   It  is  impossible  to check this based on the available results.   Likewise, in AGC\,731921 a star-forming region or a very small  galaxy is projected onto the body of the galaxy.  It is impossible to draw any conclusions due to the small number of stars.

 Several  galaxies  (AGC\,102728,  AGC\,123352 , AGC\,229379, and AGC\,229053) have an asymmetric shape that could be explained by the interaction with neighbors, but no galaxies with similar distances are observed nearby.

\section{POSSIBLE  NEIGHBORING  GALAXIES}

Searching for neighboring galaxies seems a separate big work to us.  Therefore, we will touch on this
complex issue only briefly. All of the AGC galaxies investigated by us have low masses and can enter into galaxy groups as dwarf members.  If we assume that the  radius  of  a  galaxy  group  can  be  0.5~Mpc,  then bright  galaxies  forming  groups,  which  can  include AGC  galaxies,  should  be  searched  for  within  this radius. For a distance of 10~Mpc a radius of 0.5~Mpc corresponds  to $2.9^{\circ}$.   

For  closer  groups  this  size  is even  larger.   Dozens  of  galaxies  with  velocities  less than 1000 km s$^{-1}$ that are located within this radius around each AGC galaxy from our list can be found in the search databases (NED2\footnote{https://ned.ipac.caltech.edu}, HyperLeda3\footnote{http://leda.univ-lyon1.fr}), and they can enter into the same groups as the AGC galaxies. Almost all of these galaxies have very small sizes, and there are no distance measurements for them. Among the 18 AGC galaxies, seven lie at an angular distance less  than $20^{\circ}$ from M\,87,  which  may  be  deemed  the central galaxy of the Virgo cluster.  Therefore, galaxies  from  the  Virgo  periphery,  whose  distances  are not yet known or have been measured unreliably, fall within the neighbor search radius. To identify the actual neighbors, we cannot use the radial velocities of these galaxies to determine  their distances,  because the velocity of a galaxy can change  in a wide range due to its motion inside the cluster. 

\renewcommand{\baselinestretch}{1.0}
\begin{table}[!]\vskip 10pt
	\begin{center}
	\caption{Possible neighbors to the 18 AGC galaxies}  
	\tabcolsep=0.11cm 
	\scalebox{0.99}{
		\begin{tabular}{|c|l|c|c|c|c|c|c|} \hline 
			N & Galaxy name            & $\alpha$   &$\delta$&$v_h$&$R$       & $D$     &Distance determination\\
			&                        & ($\deg $)  &($\deg$)  &(km s$^{-1})$&(\arcmin)& (Mpc) &  method \\
			\hline
			01&AGC\,102728               &   0.088333 & 31.01056 & 566 &    &  08.84    & TRGB$^*$                 \\   
			02&AGC\,123352               &  42.147500 & 23.27278 & 467 &    &  08.47    & TRGB$^*$                 \\   
			03&AGC\,198507               & 138.855833 & 25.41972 & 502 &    &  11.85    & TRGB$^*$                  \\  
			&SDSS\,J091815.92+260841.2 & 139.566370 & 26.14481 & 515 &  58&  $-$      &                       \\
			04&AGC\,198508               & 140.739583 & 24.94750 & 519 &    &  09.97    & TRGB$^*$                  \\  
			05&AGC\,198691               & 145.888750 & 33.45333 & 514 &    &  08.88    & TRGB$^*$                  \\    
			&UGC\,05186                & 145.753333 & 33.26306 & 549 &  13&  8.31     & TF [1]                \\   
			06&AGC\,200232               & 154.357917 & 29.36694 & 450 &    &  10.06    & TRGB$^*$                  \\  
			&SDSS\,J101902.38+284321.5 & 154.759941 & 28.72267 & 305 &  44&   $-$     &                       \\  
			07&AGC\,205590               & 150.144167 & 30.53917 & 494 &    &  09.34    & TRGB$^*$                  \\  
			&SDSS\,J095935.89+304845.5 & 149.899577 & 30.81266 & 651 &  21&  $-$      &                      \\  
			&UGC\,5340(DDO68)          & 149.195417 & 28.82556 & 507 & 114&  12.00    & TRGB [2]             \\ 
			&                         &            &          &     &    &  12.10    & TRGB [3]             \\ 
			&                         &            &          &     &    &  12.80    & TRGB [4]             \\  
			&UGC\,5427                 & 151.168750 & 29.36389 & 494 &  88&  11.29    & TRGB [5]             \\  
			&                         &            &          &     &    &   7.69    & TRGB [6]             \\
			&                         &            &          &     &    &   7.11    & BS [7]               \\
			&UGC\,5272                 & 147.595000 & 31.48583 & 520 & 143&   7.11    & BS [7]                \\  
			&                         &            &          &     &    &   3.80    & BS [8]                \\  
			&                         &            &          &     &    &   6.50    & TF [9]                \\  
			08&AGC\,223231               & 185.719583 & 33.83111 & 571 &    &  07.13    & TRGB$^*$                 \\  
			&UGC\,7427                 & 185.477917 & 35.05056 & 725 & 74 &   $-$     &                      \\
			09&AGC\,223254               & 187.022083 & 22.28889 & 603 &    &  06.15    & TRGB$^*$                   \\
			&UGC\,7584                 & 187.017083 & 22.58694 & 602 & 18 &  9.20     & TF [1]                 \\
			&                         &            &          &     &    &  9.95     & TF [6]                 \\
			&NGC\,4455                 & 187.185417 & 22.82167 & 643 & 34 & 6.70--12.50&  TF [1,6,9-16]       \\
			10&AGC\,229053               & 184.563750 & 25.57139 & 425 &    & 06.40     &  TRGB$^*$                   \\
			&AGC\,229100               & 185.12 9150 & 25.37056 & 221 & 33 &  $-$     &                       \\
			&SDSS\,J121531.12+253944.4 & 183.879686 & 25.66236 & 226 & 37 &  $-$      &                         \\
			&SDSS\,J121934.24+262531.5 & 184.892677 & 26.42542 & 242 & 54 &  $-$      &                          \\
			11&AGC\,229379               & 187.662917 & 23.20000 & 624 &    & 11.02     &  TRGB$^*$                   \\
			&NGC\,4455                 & 187.185417 & 22.82167 & 643 & 34 &  6.70--12.50&  TF [1,6,9-16]       \\
			&UGC\,7584                 & 187.017083 & 22.58694 & 602 & 18 &   9.20    &  TF [1]             \\
			&                         &            &          &     &    &  9.95     &  TF [6]             \\ 
			12&AGC\,238890               & 203.134583 & 25.11417 & 360 &    & 05.08     &  TRGB$^*$             \\
			&SDSS\,J133130.60+242313.3 & 202.877519 & 24.38705 & 335 & 46 &  $-$      &                  \\
			&SDSS\,J132959.46+243140.9 & 202.497765 & 24.52804 & 227 & 49 &  $-$      &                 \\
			&UGC\,8638                 & 204.834167 & 24.77000 & 274 & 95 &  4.03     & TRGB [5]       \\
			&                         &            &          &     &    &  4.29     & TRGB [15]      \\
			&                         &            &          &     &    &  4.29     & TRGB [6,18]   \\
			&                         &            &          &     &    &  2.30     & BS [19]       \\
			13&AGC\,731448               & 155.938750 & 27.11806 & 517 &    &  09.73    & TRGB$^*$        \\
			&SDSS\,J102746.49+272030.9 & 156.943724 & 27.34195 & 377 & 55 &   $-$     &          \\
			14&AGC\,731921               & 181.386250 & 28.23250 & 505 &    &  09.89    & TRGB$^*$         \\
			&AGC\,220071                & 181.350833 & 28.36750 & 565 &  8 &  $-$      &        \\
			15&AGC\,739005               & 138.409583 & 19.61889 & 429 &    &  07.83  & TRGB$^*$         \\
			&2MASS\,J09124191+1928561  & 138.174618 & 19.48237 & 348 & 16 &   $-$   &              \\
			&SDSS\,J091558.74+193914.1 & 138.994769 & 19.65395 & 377 & 33 &   $-$   &              \\
			&SDSS\,J091056.45+194931.9 & 137.735219 & 19.82554 & 342 & 40 &   $-$   &             \\
			16&AGC\,740112               & 162.477083 & 23.09000 & 609 &    &  09.90  &  TRGB$^*$       \\
			&SDSS\,J104825.55+232323.3 & 162.106467 & 23.38982 & 796 & 28 &  $-$    &            \\
			&SDSS\,J105230.99+230005.0 & 163.129177 & 23.00141 & 783 & 36 &  $-$    &            \\
			&NGC\,3344                  & 160.879167 & 24.92056 & 588 &141 &   9.82   & TRGB [6]   \\
			&                         &            &          &     &    &6.10--9.91& TF  [9,20]   \\
			17&AGC\,742601                & 192.400833 & 21.91806 & 539 &    &  06.31   &  TRGB$^*$       \\
			&IC\,3840                  & 192.942362 & 21.73640 & 583 & 32 &    5.50  &  TF  [1]    \\
			&UGC\,08011                & 193.096250 & 21.63056 & 765 & 42 & 21.40    &  TF [9]\\
			18&AGC\,747826                & 181.965833 & 31.55444 & 558 &    & 12.01    &  TRGB$^*$       \\
			&SDSS\,J120634.52+312034.7 & 181.64 3833 &31.34297 & 568 & 20 &  $-$&         \\
			&SDSS\,J120531.04+310434.1 & 181.379354 & 31.07615 & 569 & 41 &  $-$   &         \\
			&NGC\,4062                 & 181.021250 & 31.90028 & 766 & 52 &9.7--23.0  & TF [6,9,10,12,13,20-26] \\
			&IC\,2992                  & 181.316250 & 30.85306 & 611 & 53 &   12.7    & TF [1]     \\
			\hline
	\end{tabular}}
	\renewcommand{\baselinestretch}{1.0}
	\footnotesize 
	\begin{list}{}{
			\setlength\leftmargin{4mm} \setlength\topsep{2mm}
			\setlength\parsep{0mm} \setlength\itemsep{2mm} }
		\item \, The last column lists the object distance determination methods taken from NED: TRGB -- from the tip of the red giant branch, TF -- the Tully-Fisher method, BS -- from the brightest stars. The coordinates and velocities were taken from the HyperLeda database. For most objects they were determined based on the ALFALFA survey \citet{Haynes_etal_2018}. In the absence of an object in the HyperLeda database, we used data from NED. \\[-7pt]
		\item \, $^*$ -- The distance was determined in this paper.\\[-7pt]
		\item \, [1] -- \citet{Karachentsev_etal_2013}; [2] -- \citet{Tikhonov_etal_2014}; [3] -- \citet{Sacchi_etal_2016}; [4] -- \citet{Makarov_etal_2017}; [5] -- \citet{Tikhonov_2018}; [6] -- \citet{Tully_etal_2013}; [7] -- \citet{Makarova_and_Karachentsev_1998}; [8] -- \citet{Schulte-Ladbeck_and_Hopp_1998}; [9] -- \citet{Tully_and_Fisher_1988}; [10] -- \citet{Tully_etal_2016}; [11] -- \citet{Springob_etal_2009}; [12] --  \citet{Bottinelli_etal_1985}; [13] -- \citet{Willick_etal_1997}; [14] -- \citet{Yasuda_etal_1997}; [15] -- \citet{ Tully_etal_2009}; [16] -- \citet{Nasonova_etal_2011}; [17] -- \citet{Karachentsev_etal_2006};[18] -- \citet{Jacobs_etal_2009}; [19] -- \citet{Makarova_etal_1998}; [20] -- \citet{Bottinelli_etal_1984}; [21] -- \citet{Aaronson_and_Mould_1983}; [22] -- \citet{Tully_etal_1992}; [23] -- \citet{Vaucouleurs_etal_1981}; [24] -- \citet{Sorce_etal_2014}; [25] -- \citet{Theureau_etal_2007}; [26] -- \citet{Ekholm_etal_2000}. \\[-7pt]     
	\end{list} 
\label{Tab2} 
\end{center}                                                       
\end{table}

\citet{Adams_etal_2015}  searched  for  the neighbors  near similar AGC galaxies by comparing the radial velocities from the ALFALFA survey. They found the dwarf galaxy  AGC\,226967  to  enter  into  a  system  of  the same dwarf galaxies, AGC\,229490 and AGC\,229491, which  closely  resembles  the  system  AGC\,198507 from the list of our galaxies.  There are more massive neighbors  with  similar  radial  velocities  near  several AGC  galaxies. These  galaxies  together  with  the AGC  galaxies  are  probably  members  of  more  extended groups. The results of the search for neighbors to  AGC  galaxies  are  presented  in  Table~\ref{Tab2},  where $v_h$ are the heliocentric velocities of the AGC galaxies  taken  from  HyperLeda, $D$ is the  distance  to  the galaxy,  and $R$ is  the  angular  distance  to  the  neighboring galaxy.

\section{CONCLUSIONS}

 Based  on  HST images for 18 dwarf galaxies, we constructed the CM diagrams on which both
 young stars  (blue  and  red  supergiants)  and  an  old  stellar population (red giants) are seen.  For each galaxy we determined  the  position  of  the  tip  of  the  red  giant branch  (TRGB  jump)  and  the  color  index  of  the RGB.  This  allowed  us  to  determine  the  distances to   the  galaxies  and   the  metallicity  of  red   giants in  these  galaxies  based  on  the  equations  from  \citet{Lee_etal_1993}. AGC\,102728,   AGC\,198691,   and AGC\,223231  have  a  very  low  metallicity,  with  an extremely  low  metal  content  in  one  of  the  galaxies (AGC\,198691) that was determined during spectroscopic observations \citep{Hirschauer_etal_2016}. 

Star formation processes with different intensities and spatial concentrations of young stars proceed in all  galaxies. In  most  cases,  young  stars  are  distributed  over  the  galaxy  body,  but  in  some  galaxies young  stars  are  concentrated  in  small  star-forming regions.  AGC\,198507 and AGC\,739005 turned out to be binary galaxies, but this result was obtained only due to the neighbors that fell into the HST images being located very closely. It would be impossible to find such galaxies if they were outside the image. The apparent asymmetry of the galaxies can be explained by their interaction with neighbors, but, in many cases, we  do  not  know  any  neighboring  galaxies.   Since many  faint  dwarf  galaxies  with  unknown  distances are observed around the galaxies investigated by us, it is possible that close neighbors for the AGC galaxies from our list will be found while obtaining new measurements.

\begin{acknowledgements}

This   work   is   based   on   observations   with   the NASA/ESA  Hubble  Space  Telescope,  obtained  at the   Space   Telescope   Science   Institute,   which   is operated by AURA, Inc.  under contract no. NAS5--26555. These   observations   are   associated   with proposals  13750  and  15243.   In  this  paper  we  used the NED and HyperLeda databases.  We are grateful to  the  referees  for  their  useful  remarks  that  allowed the first version of the paper to be improved.

The study was financially supported by the Russian  Foundation  for  Basic  Research  and  the  National  Science  Foundation  of  Bulgaria  within  Research Project no. 19--52--18007.
\end{acknowledgements}

\end{document}